# Design of broadband optical gain in GaSb-based waveguide amplifiers with asymmetric quantum wells

IFTE KHAIRUL ALAM BHUIYAN,* JOONAS HILSKA, MARKUS PEIL, JUKKA VIHERIÄLÄ, AND MIRCEA GUINA

*Optoelectronics Research Centre, Tampere University, Korkeakoulunkatu 10, 33720 Tampere, Finland*
*ifte.alambhuiyan@tuni.fi*

**Abstract:** A design strategy for achieving broadband optical gain in GaSb-based semiconductor amplifiers operating beyond 2 µm is presented. By employing asymmetric GaInSb/AlGaAsSb quantum wells (QWs) of varying thicknesses, a flat and wide gain spectrum is demonstrated. The approach leverages carrier density and transition energy tuning across QWs to access various energy levels at specific current densities. Simulations using "Harold" self-consistent environment predict a full-width at half-maximum (FWHM) gain bandwidth exceeding 340 nm for a structure comprising one 7 nm and three 13 nm-thick QWs. The modelling parameters were validated against experimental data, ensuring a robust framework for designing broadband amplifiers and superluminescent diodes for mid-infrared applications.

1. **Introduction**

The development of optoelectronic devices operating at 2–3 µm spectral window has gained significant attention owing to an increasing number of sensing applications exploiting absorption-based molecular spectroscopy [1–4]. Moreover, if available with required performance, broadband light sources at this wavelength region could enable expansion of advanced imaging using optical coherence tomography (OCT) to new application areas, benefiting from reduced scattering at > 2 µm, and higher axial resolution linked to broadband emission [5–7]. At the same time, the 2 µm wavelength region aligns well with the transmission capabilities in thulium-doped fiber, potentially contributing to the development of free-space communication and LIDAR applications [8–10]. Finally, the recent progress of mid-infrared photonic integration technologies offers pathways to scale the production of these customized broadband devices with advanced functionalities integrated on-chip [11–14]. In general, these application areas would benefit from the availability of light sources that combine broad spectral coverage with high output power and electrical pumping. For example, a flat gain spectrum is highly desired for broadband tunable lasers allowing to decrease the variations in output power across the tuning range, or for broadband superluminescent diodes used in multi-wavelength spectroscopy and OCT imaging.

To this end, we exploit the versatility of bandgap engineering and investigate the use of asymmetric QWs to produce broadband gain profiles in GaSb-based heterostructures. Prior work at shorter wavelength regions, based on more conventional GaAs- and InP-based systems, has demonstrated the effectiveness of using bandgap engineering, strain engineering, and the variation of QW thicknesses to achieve broadened emission profiles [15–23]. Extending these design principles into the mid-infrared, specifically using GaSb-based material system, opens up compelling opportunities. Namely, GaInSb/AlGaAsSb QWs provide direct bandgap transitions with high radiative efficiency and flexibility for tuning the band structure through alloy composition and strain management [24, 25]. By varying the number of QWs, their thicknesses, and the surrounding barrier layers, it is possible to access various energy levels of confined states such that multiple energy transitions contribute to broadening the gain spectrum. However, given the large number of degrees of freedom in such structures, the design of broadband optical amplifiers requires systematic approach to optimize the gain features while taking into account physical design constraints. To this end, in this study we make use of advanced simulation tools (Harold; Photon Design [26]), taking into account carrier transport, and waveguiding optical confinement.

A simplified schematic of the simulated heterostructures used in the study is shown in Fig. 1. We first explore the simulation results with respect to single QW (SQW) structures with different thicknesses to observe basic trends setting the base of comparison with more advanced structures. Then these simulations are extended to double-QW (DQW) and multi-QW (MQW) configurations to evaluate the role of QW thickness variation in broadening the gain profile. Although it is somewhat marginal for the ultimate scope, our choice for engineering the bandgap based on changing the QW thickness instead of alloy composition is motivated by a higher accuracy in fabrication , i.e. higher precision in controlling the thickness.

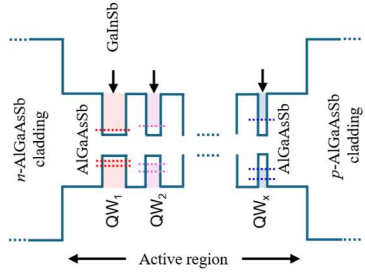

Fig. 1. Schematic of InGaSb/AlGaAsSb QW gain structure defining the generic design framework for simulation. The number of QWs and their individual thicknesses are varied during the optimization routine.

We would like to note that previous reports on using self-consistent simulations employing Harold simulator for heterostructures design operating at 2 $\mu$m wavelength range were confined to simulation of laser diodes, based on InP heterostructure operating at 2 $\mu$m [27], and GaSb heterostructures operating at 2.3 $\mu$m [28, 29].

## 2. Methodology

The Harold simulation framework employs a self-consistent coupled-physics approach, solving Poisson's equation for electrostatic potential, current continuity equations for carrier transport, capture–escape balance equations for QW carrier dynamics, photon rate equations for optical field evolution as well as heat flow equations for thermal effects. The simulation tool computes the optical mode profile by solving the vertical and longitudinal wave equations, while the Schrödinger equation is applied to determine confined states in the QWs. The optical gain is evaluated using parabolic band approximation as described by Asada et al. [30]. The model accounts for carrier transport in MQW systems, resolving the spatial distributions of electron and hole densities for both confined (in QW) and unconfined (barrier/separate confinement heterostructure) carriers. A simplifying assumption is adopted, wherein carrier capture and escape times are treated as constants across all QWs, neglecting potential variations due to local strain or compositional fluctuations.

The bandgaps and general material parameters for the $Ga_{1-x}In_xSb$ and $Al_xGa_{1-x}As_ySb_{1-y}$ materials used in the heterostructures were taken from literature [31, 32]. We simulate the gain structure in isothermal operation mode (pulsed mode). The temperature dependency of the compound semiconductor band gaps were interpolated from the binary alloy parameters using the semi-empirical Varshni model [33].

The optical gain spectra of a single QW is evaluated by the simulator following the standard expression:

$$g(E_{phot.}) = \frac{\pi\hbar q^2}{ncm_e\varepsilon_0}M^2\frac{1}{E}\sum_{j=1}K_{ij}A_{ij}(E_{phot.})D_{ij}[F_j(E_{phot.}) - F_i(E_{phot.})] \quad (1)$$

where $E_{phot.}$ is the photon energy, $K_{ij}$ is the overlap integral of the transition state $i$ and $j$, $A_{ij}$ is the dipole moment enhancement factor, $n$ is the real part of the refractive index, $q$ is the free electron charge, $m_e$ is the electron mass, $\hbar$ is the plank's constant, $\varepsilon_0$ is the free space dielectric constant, c is the speed of light, $M^2$ is the dipole moment of the bulk material, $F_j(E_{phot.})$ and $F_i(E_{phot.})$ are fermi occupation factors, $D_{ij}$ is the reduced density of states for the transition between $j$-th and $i$-th energy levels. Then the spontaneous emission rate is evaluated by

$$r_{sp}(E_{phot.}) = \frac{q^2 n^3 E_{phot.}}{\pi \hbar^2 m_e^2 c^3 \varepsilon_0} M^2 \sum_{j=1} K_{ij} D_{ij} F_j(E_{phot.})[1 - F_i(E_{phot.})] \quad (2)$$

and is consistent with the expression of gain model. The summation operation loop flows for all quantum confined states. Thus, the gain and the spontaneous emission spectrum are subsequently calculated using the following expressions:

$$G(\hbar\omega) = \int g(E_{phot.}) L(\hbar\omega - E_{phot.}) dE_{phot.} \quad (3)$$

$$R_{QW}^{sp}(\hbar\omega) = \int r_{sp}(E_{phot.}) L(\hbar\omega - E_{phot.}) dE_{phot.} \quad (4)$$

where $L(x)$ is the Lorentzian shape function. Broadening in gain and spontaneous emission expressions are based on convolution of gain and spontaneous emission rate with Lorentzian function, respectively.

Initially we designed a set of gain structures based on SQW and DQW. All active region of the gain structures included type-I compressively strained $Ga_{0.73}In_{0.27}Sb$ QWs, separated by 20 nm of $Al_{0.25}Ga_{0.75}As_{0.02}Sb_{0.98}$ barriers lattice matched to GaSb. The QWs were embedded within 130 nm waveguide material from both sides. The waveguide region was surrounded by ~1 μm thick n- and p-doped cladding layers consisting of higher band gap $Al_{0.5}Ga_{0.5}As_{0.04}Sb_{0.96}$ alloy, which is also lattice matched to GaSb. The epitaxial structure was finalized by a highly doped p-GaSb contact (cap) layer. The generic epitaxial structure is shown in Fig. 2; we show here only single QW on the figure. For double or multiple QW cases analyzed in the next sections, we used a 20 nm- $Al_{0.25}Ga_{0.75}As_{0.02}Sb_{0.98}$ barrier between adjacent QWs. The epitaxial structure was processed to define waveguide device configured as a broad-area (BA) laser diode (LD) structure with a stripe width of 80 μm and a cavity length of 2 mm.

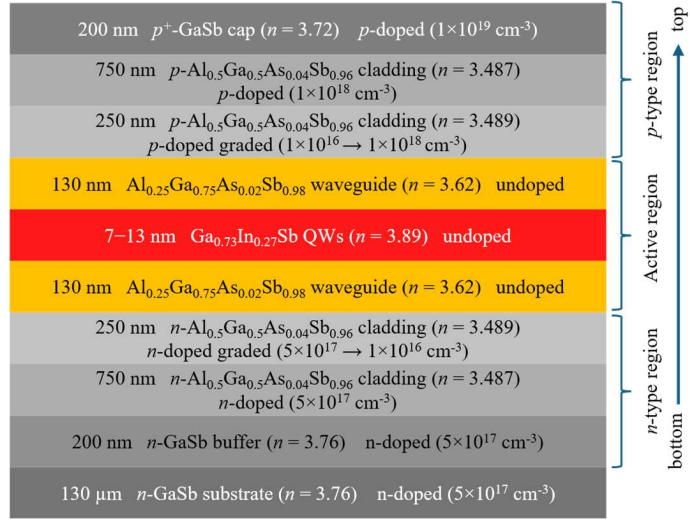

Fig. 2. Epitaxial layer structure and the key parameters considered in the simulated gain structures. The QW thickness varied from 7 nm to 13 nm. In case of double or multiple QWs, a 20 nm barrier with similar composition of waveguide was used between two adjacent QWs.

In addition to simulating the lasing action, Harold allows evaluation of QW material gain and spontaneous emission characteristics by setting the reflectivities at both mirror facets close to zero (~$10^{-7}$). To validate the modeling parameters, we initially simulated the spontaneous emission and lasing characteristics of a structure incorporating a DQW gain region. The selected design was experimentally realized, and its optical properties were characterized through photoluminescence (PL) and LD characteristic measurements, which were then used as benchmark to adjust the key parameters of the simulator. Fig. 3(a) shows the comparison fit for PL revealing an excellent matching in spontaneous emission peaks at ~1950 nm and ~2090 nm between the simulated and experimental structure. The calculated emission bandwidth is ~300 nm FWHM at a carrier density of ~3e17 $cm^{-3}$. Intensity mismatch on the shorter wavelength side is mainly due to atmospheric absorption, while minor discrepancies arise from experimental conditions (532 nm CW excitation at ~90 mW). To further validate the model, we simulated the performance of a broad-area laser diode (2 mm × 80 µm) incorporating the same DQW active region in heterostructure with slightly thicker cladding. Fig. 3(b) compares the simulated and experimental light–current (L–I) characteristics under pulsed operation at room temperature. The inset in Fig. 3(b) presents the corresponding current–voltage (I–V) characteristics. The simulation shows again good agreement with the experimental threshold behavior and output power slope across the full bias range. These results support the validity of the simulation setting the relevant parameters used in simulation. The numerical parameters used in the simulation are listed in the Table. 1.

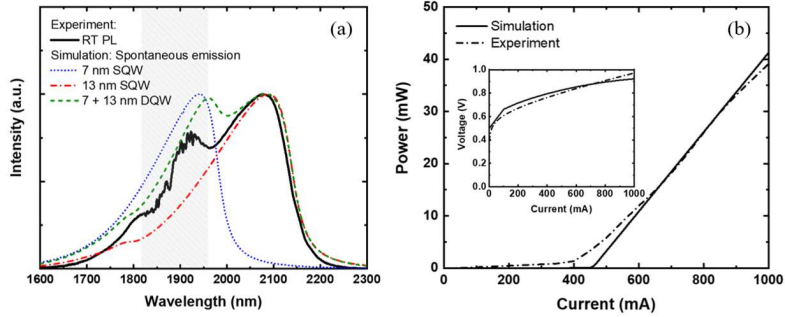

Fig. 3. (a) Room-temperature PL spectrum from a DQW gain structure with 7 nm and 13 nm wells. Simulated PL (green, short-dashed) is fitted for the same structure for carrier density ~3e17 $cm^{-3}$. Simulated gain spectra from 7 nm (blue, short-dotted) and 13 nm (red, dash-dotted) SQWs are also shown for comparison. (b) Experimental and simulated characteristics for the BA-LD structure with the same DQW heterostructure under pulsed operation.

Table. I. Simulation parameters and their values

| Parameter (symbol) | Value (unit) |
| --- | --- |
| Intraband relaxation time ($\tau_{in}$) | 1e-13 s |
| Shockley-Read-Hall lifetime of electron ($\tau_{e-SRH}$) | 1e-07 s |
| Shockley-Read-Hall lifetime of hole ($\tau_{h-SRH}$) | 1e-07 s |
| Capture time constant of electron ($\tau_e^{cap}$) | 5e-12 s |
| Capture time constant of hole ($\tau_h^{cap}$) | 3e-13 s |
| Gain factor | 0.6 |
| Fraction of spontaneous emission coupled into optical mode | 0.0001 |
| Mirror reflectivity (both facets) | 1e-07 |
| Scattering loss ($cm^{-1}$) | 4.5 |

## 3. Design and optimization of broadband gain structure with SQW and MQWs

We first simulated gain structures comprising SQW with different thickness of the QW, i.e, from 7 nm to 13 nm with one nm intervals. Then in a second iteration we simulated DQW structures with asymmetric QWs wherein the top QW thickness (see in Fig. 4) was fixed to 7 nm and the bottom QW varied in thickness from 7 nm to 13 nm, again with one nm intervals. A key aspect of the optimization was to understand the contribution for individual QWs to the gain spectrum, and their combined contributions in a DQW structure, and corresponding dependence on carrier densities. Fig. 4 summarizes the exemplary simulated gain curves at 20°C for SQW and DQW structures. The simulation reveal important distinctive features in the wavelength variation with carrier density, with clear transitions form gain peak corresponding to E1-HH1 QW transition to E1-LH1 transitions when carrier density increases.

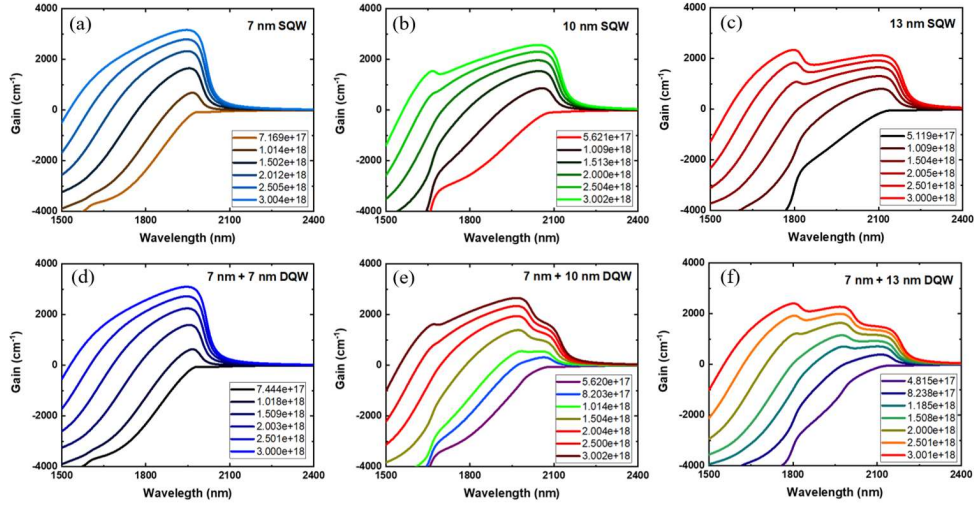

Fig. 4. Exemplary gain spectrum of different (a-c) SQWs of thickness 7 nm, 10 nm and 13 nm, and (d-f) DQWs with top QW thickness fixed to 7 nm, and bottom QW thickness of 7 nm, 10 nm, and 13 nm. All legends in the figures show the carrier density level in the QWs.

Fig. 5 summarizes the gain spectrum maxima (gain peak wavelength) variation with the carrier density. Up to a carrier density in the range of ~2.5e18 cm$^{-3}$ (corresponding to a current injection level of ~1600 mA), the gain is dominated by the E1-HH1 transition for all SQW structures (see in Fig 5(a)). Beyond this carrier density, significant contribution from the E1-LH1 transition appears. For example, in 13 nm QW, a sharp shift in the gain peak from 2097 nm to 1800 nm at ~2.6e18 cm$^{-3}$ indicates the onset of broader gain spectra. Similar transition are observed in the 10 nm, 11 nm, and 12 nm SQWs at carrier densities exceeding ~3e18 cm$^{-3}$.

For the DQW gain structures (see in Fig. 5(b)), the simulations showed that at a carrier density of ~1.2e18 cm$^{-3}$ both QWs exhibit gain maxima corresponding to E1-HH1 transition, which varies with QW thickness. Moreover, the DQW structures containing bottom QWs with thickness of 12 nm and 13 nm QWs showed clear E1-LH1 transition at ~1967 nm for a carrier density of ~3.4e18 cm$^{-3}$ and ~2.6e18 cm$^{-3}$, respectively. This E1-LH1 transition could not be seen for DQW structures with thinner "bottom" QWs, since we limit our simulation to injection current level of 10 A (corresponds to a carrier density of ~4.5e18 cm$^{-3}$); observation of the thinner QWs E1-LH1 transition would require higher carrier density, which is considered impractical. In general, our analysis shows that at a specific carrier density, a balanced gain condition can be achieved, wherein transitions from both QWs add to form a broad and flat gain spectrum. However, under higher injection levels, carrier occupation becomes higher in the thinner QW with larger density of states, resulting in a spectral shift of peak emission towards that QW.

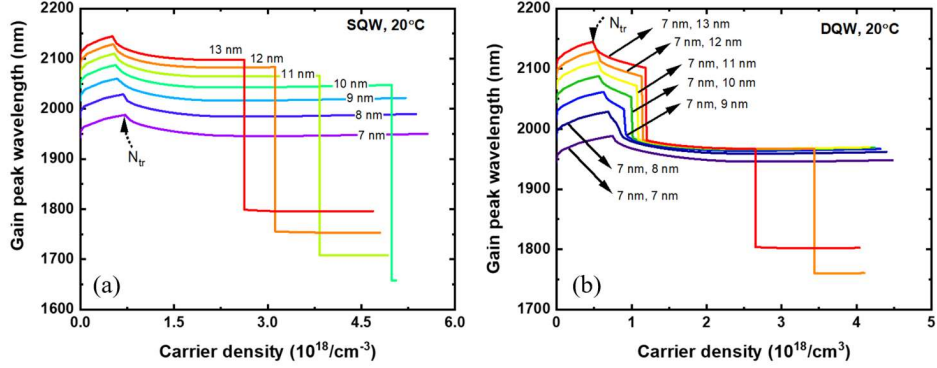

Fig. 5. Evolution of the QW gain peak wavelength as a function of carrier densities for (a) SQWs of different thicknesses varying from 7 nm to 13 nm. (b) Similar variation is shown for DQWs with the QW near the p-cladding is fixed to 7 nm, and the thickness of other QW is varied from 7 to 13 nm. In both cases $N_{tr}$ specifies transparency carrier density level.

Based on the observations made, several combinations of thicknesses employed in DQW heterostructures could be found for which wide and flat gain profile was achieved at specific carrier densities. For example, Fig. 6 shows simulated gain spectra having flat gain and wide spectral coverage for a DQW structure with fixed top QW of 7 nm, and bottom QW thickness varied from 9 nm to 13 nm. The peak gain corresponding to the 7 nm QW is always near ~1980 nm, while the gain peak for thicker QW changes from ~2030 nm to ~2100 nm with the thickness. A maximum gain bandwidth of ~256 nm FWHM can be achieved for a combination of 7 nm and 13 nm QWs at higher carrier density of N=1.2e18 cm$^{-3}$ (corresponds to ~340 mA current injection). In this case a maximum gain of ~710 cm$^{-1}$ is maintained between 1953 nm and 2140 nm, with a variation of ~70 cm$^{-1}$ in the overlapping region of the different contributions to the total gain spectra.

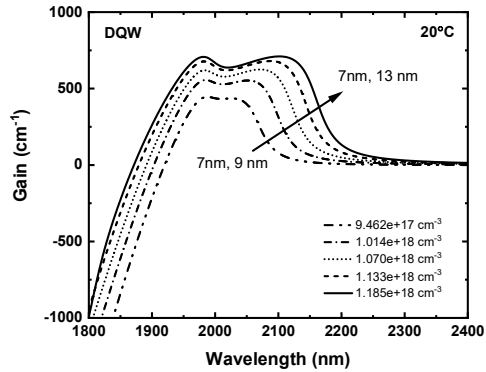

Fig. 6. Flat gain spectrum obtained for asymmetric DQW structures employing a fixed bottom QW with a thickness of 7 nm and a top QW with varying thickness from 10 nm to 13 nm (varying in 1 nm intervals in increasing order as highlighted by the arrow). The legend denotes the carrier density (cm$^{-3}$) levels for this simulation.

To further extend the spectral bandwidth and emission power, we simulated MQW devices with more than two QWs, while keeping the total QW thickness below strain relaxation. More specifically, we set a total strain-times-thickness limit of around 100 %×nm as a rough estimate based on available data for GaInSb [25] and AlGaAsSb [34] strain relaxation when grown on GaSb substrate. Within these constraints, the MQW structures were systematically optimized to achieve a balance between spectral flatness and bandwidth (in FWHM) of the gain spectrum at high carrier density.

Fig. 7 shows flat gain spectra at RT from three different structures: A, B and C, comprising two different types of QW thickness (7 nm and, 13 nm) with uniform composition. Structure A corresponds to the DQW structure analyzed in Fig. 3 and acts as the reference. Structure B includes one additional 13 nm QW compared to structure A and shows an improved flat gain level from 706 cm$^{-1}$ to 1032 cm$^{-1}$ at an elevated carrier density. Moreover, structure C that comprising one 7 nm and three 13 nm QWs produces a yet broader gain profile, at a higher carrier density level compared to other structures. We limit inclusion of additional QWs in the simulation to not exceed the critical strain limit in the active region. We note employing strain-compensation architectures [35, 36] could be utilized for further optimization however that is considered out-of-scope for this work.

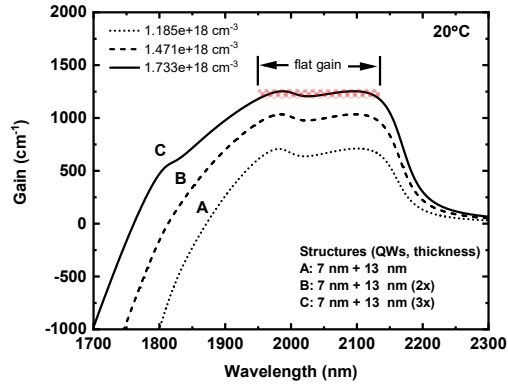

Fig. 7. Simulated gain spectra for various MQW designs.

The simulation shows that the thin QW contributes to gain at shorter wavelengths (around ~1980 nm), while the thicker QWs dominate in the longer-wavelength region (around ~2100 nm) at a carrier density level of ~1.73e18 cm$^{-3}$ (corresponds to ~2.1 A injection current). Maximum flat gain level (red shaded region in gain spectra of Fig. 7) for structure C covers a wavelength range ~185 nm (from ~1950 nm to ~2135 nm) with a gain fluctuation of 73 cm$^{-1}$. The calculated maximum gain is ~1253 cm$^{-1}$, which is approximately two-fold improvement compared to reference gain structure A. The gain spectrum for structure C shows double gain peaks around ~1987 nm and ~2096 nm due to the combined energy transition from the four QWs. The maximum bandwidth of the gain spectra from the optimized structure C has a FWHM of ~342 nm, which is approximately 100 nm broader compared to structure A.

Fig. 8(a) shows simulation of RT gain spectra at different carrier densities for structure C. At a moderate carrier density level ~1e18 cm$^{-3}$, peak gain appears near ~2100 nm only due to dominant carrier transition from 13 nm QWs. As mentioned earlier, flat and balanced gain spectra appear around ~1.73e18 cm$^{-3}$ having distinguishable double peaks with a relative small dip in between. However, at a higher carrier density level of ~2.14e18 cm$^{-3}$, the gain extends to shorter wavelength region around ~1800 nm with a degraded flatness. The third peak around ~1800 nm is generated due to excitation from higher energy levels favoring the thicker QW.

Fig. 8(b) shows the gain profiles for structure C# but for a shorter device. Here the length of the active waveguide (current injection region) was reduced from 2 mm to 1775 μm to, while 225 μm of the waveguide end was not pumped to implement a mirror loss of ~95 cm$^{-1}$ without modifying the facet reflectivity. In practice, this strategy allows us to suppress cavity feedback while preventing lasing and allows us to use higher injection densities to access E1-LH1 transition. In this case, the gain enhancement is primarily attributed to the contribution of the second quantized state originating from the thicker quantum well (QW), at a carrier density of N=2.412e18 cm$^{-3}$ (corresponds to ~4.4 A). We would note that Mehuys *et al.* [37] employed a similar strategy to enhance gain bandwidth by optimizing cavity loss to enable lasing from both the n=1 and n=2 quantum states, resulting in a broadened and flattened gain spectrum. The fluctuation in the non-uniformity or flatness corresponds to QW gain variation of ~278 cm$^{-1}$ over a broad spectral window spanning 1760–2150 nm. The peak gain values are calculated as ~1600 cm$^{-1}$ at ~2100 nm (E1-HH1, 13 nm QW), ~1733 cm$^{-1}$ at 1980 nm (E1-HH1, 7 nm QW), ~1670 cm$^{-1}$ at 1800 nm (E1-LH1, 13 nm QW). This optimization yields a gain enhancement of approximately 400–500 cm$^{-1}$ for both thick and thin QWs, respectively.

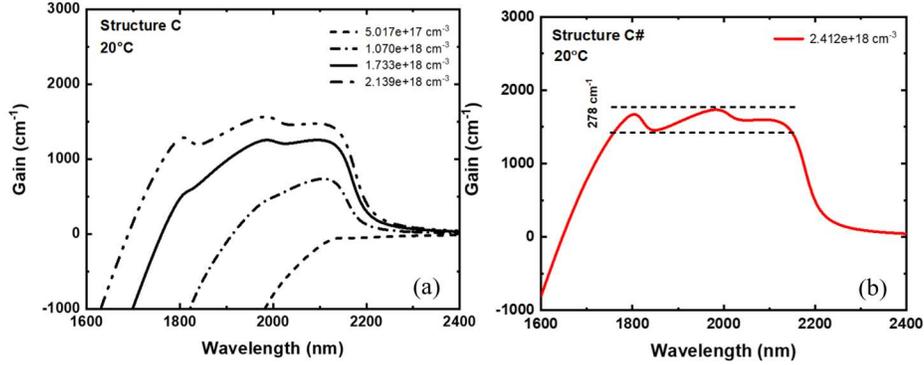

Fig. 8. (a) Simulated gain spectra for structure C at different carrier densities. (b) Gain spectra of structure C# for a modified waveguide/cavity length (1775 µm) at a specific carrier density (at N=2.412e18 cm$^{-3}$) showing elevated gain for E1-HH1 transition from all QWs as well as E1-LH1 transition from 13 nm QW.

A very important aspect for practical device operation is temperature dependencies. Fig. 9 shows simulated gain spectrum at temperatures between 10°C and 100°C, revealing the change of the carrier density required to attain a flat gain spectral characteristic. Owing to the temperature increase, the gain peak at the sorter wavelength region increased from 1970 nm to 2113 nm, while for the longer wavelength region of the gain peak increased from 2078 nm to 2237 nm. Moreover, the FWHM of the gain spectra varies from ~335 nm to a maximum ~400 nm as temperature increases. However, as expected, the maximum gain was significantly reduced (~195 cm$^{-1}$) with increasing temperature.

At an elevated injection, charge carriers tend to populate at the thinner QW due to its stronger confinement and higher density of state (DOS) at the ground state, resulting in higher optical gain relative to the thicker well. This behavior leads to spectral tilt and reduced uniformity in gain spectrum and underscores a fundamental trade-off: while increasing the number of QW can enhance overall gain and broaden the emission spectrum, it does not guarantee proportionally higher gain especially when carrier distribution becomes increasingly non-uniform across the wells. Further optimization of the gain structure to achieve better carrier uniformity may include the use of energy barriers between groups of QWs to minimize carrier leakage and improve carrier confinement. Combining the simulation with physically realizable structural design, this study shows a pathway to develop broadband gain devices in the 2 $\mu$m wavelength region and beyond. The insights herein also bridge a critical gap between understanding the theoretical potential of these novel broadband devices, while minimizing the need for extensive experimental iterations of device development.

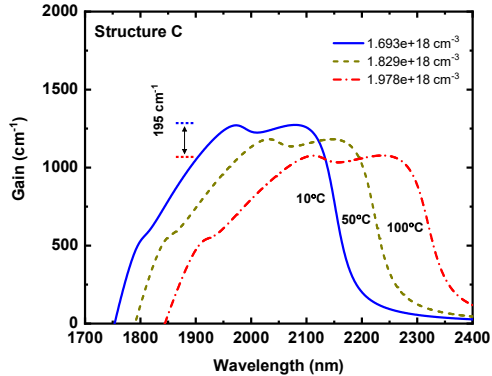

Fig. 9. Temperature dependency of the gain spectra for structure C.

## 4. Conclusion

A design strategy for achieving broadband optical gain in GaSb-based semiconductor optical amplifiers operating beyond 2 µm was proposed. Through systematic bandgap engineering, we explored the impact of varying QW thicknesses within asymmetric MQW heterostructures on the gain spectrum. To this end we used a validated self-consistent simulation with the Harold framework. Our analysis revealed that the use of asymmetric GaInSb/AlGaAsSb QWs enables broader gain spectrum by exploiting the interplay between carrier density-dependent energy transitions (corresponding to E1-HH1 and E1-LH1 transitions) in different QWs.

Simulations of both SQW and DQW structures established the relationship between carrier injection levels and the onset of multi-transition gain contributions. Notably, a MQW structure with one 7 nm and three 13 nm QW was found to deliver a flat and wide gain profile, maintaining a high gain (~1253 cm$^{-1}$) across a spectral range of ~1953–2140 nm with a full-width at FWHM exceeding 340 nm for room-temperature operation. Furthermore, the simulations revealed the effects of temperature increase, demonstrating the possibility for even broadening the gain spectrum with a FWHM of ~400 nm at ~100°C at the expense of reduced gain. The simulation results were validated against experimental data obtained from PL and LD characterization, confirming the accuracy of the modeling parameters and supporting the practicality of the proposed design approach. Future work will focus on further expanding the gain bandwidth through the inclusion of additional QWs with optimized thickness and strain profiles, aiming also at wavelength extension towards 3 µm.

**Funding**

This work was supported by the European Commission under the Horizon 2020 program through the project Next Generation of Tunable Lasers for Optical Coherence Tomography (NETLAS, Grant Agreement No. 860807, H2020-MSCA-ITN-2019). Additional support was provided by the Academy of Finland Flagship Programme PREIN (Decision No. 320168).

**Acknowledgement**

The authors acknowledge Dr. Topi Uusitalo, and Dr. Nouman Zia for fruitful discussion on the simulation.

**Disclosures**

The authors declare no conflicts of interest.

**Data availability**

Data underlying the results presented in this paper can be obtained from the authors upon reasonable request.

**References**

1. Vizbaras, K., Dvinelis, E., Šimonytė, I., Trinkūnas, A., Greibus, M., Songaila, R., Žukauskas, T., Kaušylas, M., Vizbaras, A.: High power continuous-wave GaSb-based superluminescent diodes as gain chips for widely tunable laser spectroscopy in the 1.95–2.45 µm wavelength range. Applied Physics Letters. 107, 011103 (2015). https://doi.org/10.1063/1.4926367
2. Vizbaras, A., Šimonytė, I., Miasojedovas, A., Trinkūnas, A., Bučiūnas, T., Greibus, M., Naujokaitė, G., Torcheboeuf, N., Droz, S., Boiko, D., Dambrauskas, Ž., Gulbinas, A., Vizbaras, K.: Swept-wavelength lasers based on GaSb gain-chip technology for non-invasive biomedical sensing applications in the 1.7–2.5 µm wavelength range. Biomed. Opt. Express. 9, 4834 (2018). https://doi.org/10.1364/BOE.9.004834
3. Zia, N., Viheriälä, J., Koivusalo, E., Virtanen, H., Aho, A., Suomalainen, S., Guina, M.: GaSb superluminescent diodes with broadband emission at 2.55 µm. Applied Physics Letters. 112, 051106 (2018). https://doi.org/10.1063/1.5015974
4. Ojanen, S., Viheriälä, J., Zia, N., Koivusalo, E., Hilska, J., Tuorila, H., Guina, M.: Discretely Tunable (2594, 2629, 2670 nm) GaSb/Si3N4 Hybrid Laser for Multiwavelength Spectroscopy. Laser & Photonics Reviews. 17, 2300492 (2023). https://doi.org/10.1002/lpor.202300492
5. Liang, H., Lange, R., Peric, B., Spring, M.: Optimum spectral window for imaging of art with optical coherence tomography. Applied Physics B. 111, 589–602 (2013). https://doi.org/10.1007/s00340-013-5378-5
6. Cheung, C.S., Daniel, J.M.O., Tokurakawa, M., Clarkson, W.A., Liang, H.: High resolution Fourier domain optical coherence tomography in the 2 µm wavelength range using a broadband supercontinuum source. Opt. Express. 23, 1992 (2015). https://doi.org/10.1364/OE.23.001992
7. Read, M., Cheung, C.S., Liang, H., Meek, A., Korenberg, C.: A Non-invasive Investigation of Egyptian Faience Using Long Wavelength Optical Coherence Tomography (OCT) at 2 µm. Studies in Conservation. 67, 168–175 (2022). https://doi.org/10.1080/00393630.2020.1871208
8. Li, Z., Heidt, A.M., Daniel, J.M.O., Jung, Y., Alam, S.U., Richardson, D.J.: Thulium-doped fiber amplifier for optical communications at 2 µm. Opt. Express. 21, 9289–9297 (2013). https://doi.org/10.1364/OE.21.009289
9. Flannigan, L., Yoell, L., Xu, C.: Mid-wave and long-wave infrared transmitters and detectors for optical satellite communications—a review. J. Opt. 24, 043002 (2022). https://doi.org/10.1088/2040-8986/ac56b6
10. Chin, S., Holzer, J., Groote, A.D., Martens, D., Naujokaitė, G., Vizbaras, A., Vizbaras, K., Pache, C.: Development of hybrid photonic integrated wavelength-tunable laser at 2 µm and its application to FMCW LiDAR. Opt. Express. 32, 22470 (2024). https://doi.org/10.1364/OE.522398
11. Zia, N., Ojanen, S.-P., Viheriala, J., Koivusalo, E., Hilska, J., Tuorila, H., Guina, M.: Widely tunable 2 µm hybrid laser using GaSb semiconductor optical amplifiers and a $Si_3N_4$ photonics integrated reflector. Opt. Lett. 48, 1319 (2023). https://doi.org/10.1364/OL.480867
12. Tournié, E., Monge Bartolome, L., Rio Calvo, M., Loghmari, Z., Díaz-Thomas, D.A., Teissier, R., Baranov, A.N., Cerutti, L., Rodriguez, J.-B.: Mid-infrared III–V semiconductor lasers epitaxially grown on Si substrates. Light: Science & Applications. 11, 165 (2022). https://doi.org/10.1038/s41377-022-00850-4
13. Remis, A., Monge-Bartolome, L., Paparella, M., Gilbert, A., Boissier, G., Grande, M., Blake, A., O'Faolain, L., Cerutti, L., Rodriguez, J.-B., Tournié, E.: Unlocking the monolithic integration scenario: optical coupling between GaSb diode lasers epitaxially grown on patterned Si substrates and passive SiN waveguides. Light: Science & Applications. 12, 150 (2023). https://doi.org/10.1038/s41377-023-01185-4
14. Ojanen, S., Viheriälä, J., Zia, N., Koivusalo, E., Hilska, J., Tuorila, H., Guina, M.: Widely Tunable (2.47–2.64 µm) Hybrid Laser Based on GaSb/GaInAsSb Quantum-Wells and a Low-Loss Si3N4 Photonic Integrated Circuit. Laser & Photonics Reviews. 17, 2201028 (2023). https://doi.org/10.1002/lpor.202201028
15. Mikami, O., Yasaka, H., Noguchi, Y.: Broader spectral width InGaAsP stacked active layer superluminescent diodes. (1990)
16. Ching-Fuh Lin, Bor-Lin Lee, Po-Chien Lin: Broad-band superluminescent diodes fabricated on a substrate with asymmetric dual quantum wells. IEEE Photonics Technology Letters. 8, 1456–1458 (1996). https://doi.org/10.1109/68.541548
17. Hamp, M.J., Cassidy, D.T.: Critical design parameters for engineering broadly tunable asymmetric multiple-quantum-well lasers. IEEE J. Quantum Electron. 36, 978–983 (2000). https://doi.org/10.1109/3.853559
18. Wang, J., Hamp, M.J., Cassidy, D.T.: Design Considerations for Asymmetric Multiple Quantum Well Broad Spectral Width Superluminescent Diodes. IEEE J. Quantum Electron. 44, 1256–1262 (2008). https://doi.org/10.1109/JQE.2008.2003104
19. Hamp, M.J., Cassidy, D.T., Robinson, B.J., Zhao, Q.C., Thompson, D.A.: Effect of barrier thickness on the carrier distribution in asymmetric multiple-quantum-well InGaAsP lasers. IEEE Photon. Technol. Lett. 12, 134–136 (2000). https://doi.org/10.1109/68.823494

20. Lin, C.-F., Wu, B.-R., Laih, L.-W., Shih, T.-T.: Sequence influence of nonidentical InGaAsP quantum wells on broadband characteristics of semiconductor optical amplifiers–superluminescent diodes. Opt. Lett. 26, 1099 (2001). https://doi.org/10.1364/OL.26.001099
21. Tsai, C.-W., Chang, Y.-C., Shmavonyan, G.S., Su, Y.-S., Lin, C.-F.: Extremely broadband superluminescent diodes/semiconductor optical amplifiers in optical communication band. Presented at the Integrated Optoelectronics Devices , San Jose, CA June 16 (2003)
22. Kwon, O.-K., Kim, K., Sim, E.-D., Kim, J.-H., Kim, H.-S., Oh, K.-R.: Asymmetric multiple-quantum-well laser diodes with wide and flat gain. Opt. Lett. 28, 2189 (2003). https://doi.org/10.1364/OL.28.002189
23. Ohgoh, T., Mukai, A., Yaguchi, J., Asano, H.: Demonstration of 1.0 μm InGaAs High-Power and Broad Spectral Bandwidth Superluminescent Diodes by Using Dual Quantum Well Structure. Appl. Phys. Express. 6, 014101 (2013). https://doi.org/10.7567/APEX.6.014101
24. Lazzari, J.L., Fouillant, C., Grunberg, P., Leclercq, J.L., Joullié, A., Schiller, C.: Critical layer thickness in AlGaAsSbGaSb heterostructures determined by X-ray diffraction. Journal of Crystal Growth. 130, 96–100 (1993). https://doi.org/10.1016/0022-0248(93)90840-S
25. Nilsen, T.A., Breivik, M., Selvig, E., Fimland, B.O.: Critical thickness of MBE-grown Ga1−xInxSb (x<0.2) on GaSb. Journal of Crystal Growth. 311, 1688–1691 (2009). https://doi.org/10.1016/j.jcrysgro.2008.11.083
26. HAROLD by Photon Design (2023).  [Online], https://photond.com/harold
27. Al-Muhanna, A., Salhi, A.: Numerical analysis of InGaAs–InP multiple-quantum well laser emitting at 2 μm. Opt Quant Electron. 46, 851–861 (2014). https://doi.org/10.1007/s11082-013-9796-8
28. Chenini, L., Aissat, A., Halbwax, M., Vilcot, J.P.: Performance simulation of an InGaSb/GaSb based quantum well structure for laser diode applications. Physics Letters A. 467, 128711 (2023). https://doi.org/10.1016/j.physleta.2023.128711
29. Salhi, A., Al-Muhanna, A.A.: Self-Consistent Analysis of Quantum Well Number Effects on the Performance of 2.3-μm GaSb-Based Quantum Well Laser Diodes. IEEE J. Select. Topics Quantum Electron. 15, 918–924 (2009). https://doi.org/10.1109/JSTQE.2008.2012000
30. M. Asada, A. Kameyama, Y. Suematsu: Gain and intervalence band absorption in quantum-well lasers. IEEE Journal of Quantum Electronics. 20, 745–753 (1984). https://doi.org/10.1109/JQE.1984.1072464
31. Piprek, J.: Semiconductor optoelectronic devices : introduction to physics and simulation. Academic Press, Amsterdam ; Tokyo (2003)
32. Vurgaftman, I., Meyer, J.R., Ram-Mohan, L.R.: Band parameters for III–V compound semiconductors and their alloys. Journal of Applied Physics. 89, 5815–5875 (2001). https://doi.org/10.1063/1.1368156
33. Varshni, Y.P.: Temperature dependence of the energy gap in semiconductors. Physica. 34, 149–154 (1967). https://doi.org/10.1016/0031-8914(67)90062-6
34. Lazzari, J.L., Fouillant, C., Grunberg, P., Leclercq, J.L., Joullié, A., Schiller, C.: Critical layer thickness in AlGaAsSbGaSb heterostructures determined by X-ray diffraction. Journal of Crystal Growth. 130, 96–100 (1993). https://doi.org/10.1016/0022-0248(93)90840-S
35. Li, W., Héroux, J.B., Shao, H., Wang, W.I.: Strain-compensated InGaAsSb/AlGaAsSb mid-infrared quantum-well lasers. Applied Physics Letters. 84, 2016–2018 (2004). https://doi.org/10.1063/1.1687981
36. W. Li, H. Shao, D. Moscicka, T. Unuvar, W. I. Wang: Strain-compensated InGaAsSb multiple quantum-wells with digital AlGaAsSb barriers for midinfrared lasers. IEEE Photonics Technology Letters. 17, 2274–2276 (2005). https://doi.org/10.1109/LPT.2005.857979
37. Mehuys D, Mittelstein M, Yariv A, Sarfaty R, Ungar J.E: Optimised Fabry-Perot (AlGa)As quantum-well lasers tunable over 105 nm. Electronics Letters. 25, 143–145 (1989). https://doi.org/10.1049/el:19890104